\documentclass[reprint, superscriptaddress, secnumarabic, amssymb, nobibnotes, aps, prb, showkeys]{revtex4-2}
\usepackage{graphicx}
\usepackage{epstopdf}
\usepackage{fontenc}
\usepackage{amsbsy}
\usepackage{amsfonts}
\usepackage{amsmath}
\usepackage{bbm}
\usepackage{braket}
\usepackage{xcolor}
\usepackage{textcomp}
\usepackage{gensymb} 
\usepackage{hyperref}
\allowdisplaybreaks

\hypersetup{            
    unicode=false,          
    pdftoolbar=true,        
    pdfmenubar=true,        
    pdffitwindow=false,     
    pdfstartview={FitH},    
    pdftitle={Re2Zr: Unconventional Superconductor based on a Kagome Lattice of Re},    
    pdfauthor={Manasi Mandal},      
    pdfsubject={},   
    pdfcreator={},   
    pdfproducer={}, 
    pdfkeywords={}, 
    pdfnewwindow=true,      
    colorlinks=true,       
    linkcolor=blue,         
    citecolor=blue,         
    filecolor=magenta,      
    urlcolor=blue           
}

\usepackage[normalem]{ulem}

\newcommand{\equref}[1]{Eq.~(\ref{#1})}

\newcommand{\figref}[1]{Fig.~\ref{#1}}

\newcommand{\tableref}[1]{Table~\ref{#1}}

\renewcommand{\approx}{\simeq}

\begin{document}

\title{%
  \texorpdfstring{%
    Time-Reversal Symmetry Breaking in Re-Based Kagome Lattice Superconductor%
  }{%
    Re2Zr: Unconventional Superconductor based on a Kagome Lattice of Re%
  }%
}
\author{Manasi Mandal}
\affiliation{Department of Physics, Indian Institute of Science Education and Research Bhopal, Bhopal, 462066, India}
\author{A.~Kataria}
\affiliation{Department of Physics, Indian Institute of Science Education and Research Bhopal, Bhopal, 462066, India}
\author{P. K.~Meena}
\affiliation{Department of Physics, Indian Institute of Science Education and Research Bhopal, Bhopal, 462066, India}
\author{R. K.~Kushwaha}
\affiliation{Department of Physics, Indian Institute of Science Education and Research Bhopal, Bhopal, 462066, India}
\author{D.~Singh}
\affiliation{Department of Physics, Indian Institute of Science Education and Research Bhopal, Bhopal, 462066, India}
\author{P. K.~Biswas}\thanks{Deceased}
\affiliation{ISIS Facility, STFC Rutherford Appleton Laboratory, Didcot OX11 0QX, United Kingdom}
\author{R.~Stewart}
\affiliation{ISIS Facility, STFC Rutherford Appleton Laboratory, Didcot OX11 0QX, United Kingdom}
\author{A.~D.~Hillier}
\affiliation{ISIS Facility, STFC Rutherford Appleton Laboratory, Didcot OX11 0QX, United Kingdom}
\author{R.~P.~Singh}
\email[]{rpsingh@iiserb.ac.in}
\affiliation{Department of Physics, Indian Institute of Science Education and Research Bhopal, Bhopal, 462066, India}

\begin{abstract}
We investigated the Re-based kagome superconductor Re$_2$Zr through various measurements, including resistivity, magnetization, specific heat, and muon spin rotation and relaxation spectroscopy. These results suggest that Re$_2$Zr is a moderately coupled potential two-gap superconductor. Zero-field muon relaxation data indicate the possible presence of a time-reversal symmetry-breaking state in the superconducting ground state. Our investigation identifies Re$_{2}$Zr as a new unconventional superconductor with a potential complex order parameter that warrants considerable experimental and theoretical interest.
\end{abstract}
\maketitle
\section{INTRODUCTION}
Understanding unconventional superconductivity is a key focus in condensed matter physics, as the mechanism behind electron pairing remains unresolved. The crystal structure of a material significantly affects its electronic and superconducting properties. Recently, frustrated structures, such as kagome lattices, have attracted considerable attention due to their potential to host unconventional superconducting ground states \cite{Kagome_sc,Kagome_sc1,Kagome_sc2,Kagome_sc3}. Kagome lattice materials exhibit various exotic electronic features, including flat bands, Dirac cones, and non-trivial topological surface states, which have been both predicted \cite{Kagome_top1,Kagome_top3} and observed \cite{Kagome_top,Kagome_top2}. The electronic correlations and intrinsic properties of the flat bands in kagome systems are believed to play a crucial role in the emergence of superconductivity, as demonstrated in the recently discovered AV$_3$Sb$_5$ compounds (where A = K, Cs, Rb) \cite{Kagome_sc,CsV3Sb5,RbV3Sb5} and intermetallic ``132" RT$_3$X$_2$ phases (where R is a rare earth metal, T is typically a 4d or 5d transition metal and X is B, Ga, or Si) \cite{LaRh3B2,lair3ga2, LaRu3Si2}. Notably, in AV$_3$Sb$_5$ systems, superconductivity is accompanied by the breaking of translational, rotational, and time-reversal symmetries \cite{Natmat,Kagome_top2}.\\

Breathing kagome metal superconductors ROs$_2$ (R = Sc, Y, Lu, Zr and Hf) have emerged as a promising platform for studying unconventional superconductivity. The kagome lattice formed by Os atoms in these compounds contributes to flat bands near the Fermi level,  which are believed to play a role in their superconducting properties \cite{ROs2, HfOs2}.  In the isostructural compound Re$_2$Hf, a time-reversal symmetry-breaking pairing state has been observed \cite{Re2Hf}, where the interplay between intra-band and inter-band Fermi surface nesting is thought to lead to unique superconducting ground states, with the superconducting gap changing sign across the nesting wavevector. Furthermore, the \textit{ab initio} band structure calculations, corroborated by experimental evidence, have revealed a novel $s+is'$ state in Re$_2$Hf \cite{Re2Hf}. Given the structural similarity between Re$_2$Hf and Re$_2$Zr, the latter compound presents an intriguing opportunity to explore unconventional superconductivity in kagome lattices. While Re$_2$Zr adopts a centrosymmetric C14 Laves phase, its underlying kagome net interleaved with triangular layers offers a unique platform for studying the interplay between kagome lattice geometry and superconducting properties.

Moreover, Re$_2$Zr can shed light on the unconventional superconducting ground state of Re-based compounds like Re$_6$X (X = Zr, Hf and Ti). These compounds, with their non-centrosymmetric $\alpha$-Mn structure, exhibit broken time-reversal symmetry (TRS), while Re-free compounds with similar structures preserve TRS \cite{ReZr,ReHf,Ti,Ti2, TaOs,NbOs,NbOs2}. This suggests that Re's local electronic structure is crucial for understanding the unconventional superconducting states in the Re$_6$X series \cite{Re}. The variable Re/X ratio in Re-X binary compounds presents a unique opportunity to explore the relationship between Re and crystal structure in Re-based unconventional superconductors, making Re$_2$Zr a promising candidate for such studies.

\begin{figure*}[ht!]
\includegraphics[width=2.05\columnwidth]{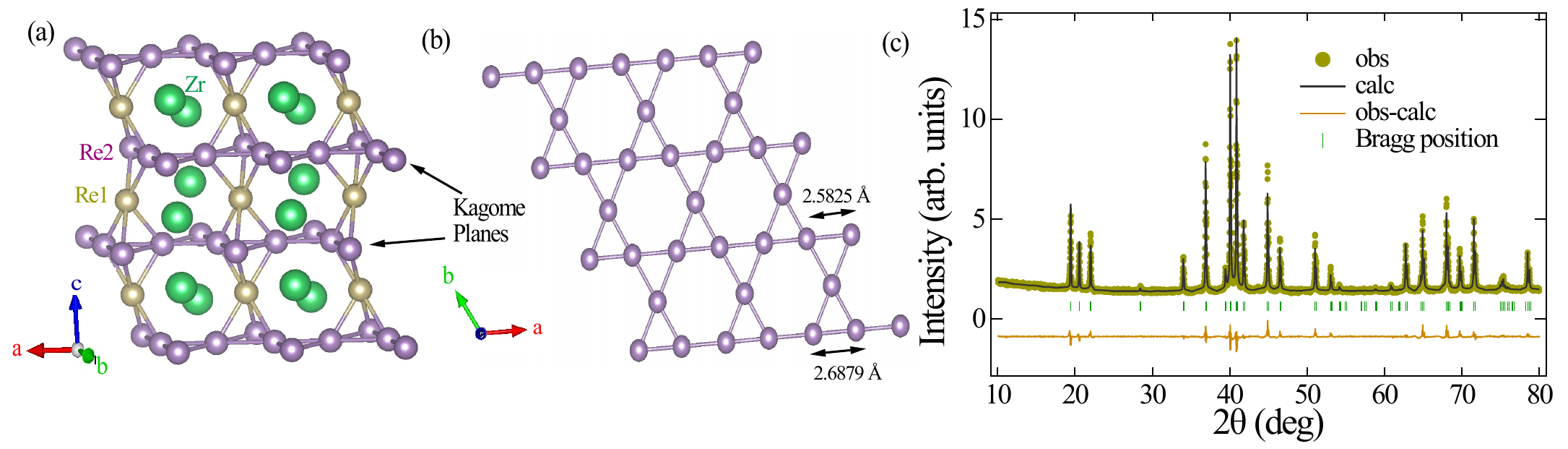}
\caption{\label{Fig1} (a) Crystal structure of Re$_2$Zr in the hexagonal MgZn$_{2}$-type (C14) structure. (b) Distorted kagome layer in the $ab$-plane, showing Re$2$ atoms forming a network where two triangles and two hexagons meet at each vertex. (c) Rietveld refined powder XRD pattern, confirming crystallization in the space group $P6_{3}$/$mmc$.}
\end{figure*}

This article presents a comprehensive study of the superconducting and normal-state properties of Re$_2$Zr. The crystal structure of Re$_2$Zr is composed of stacked kagome layers connected by intermediate 3D triangular layers, as illustrated in \figref{Fig1}(a). The Re kagome layer consists of two non equilateral corner sharing triangles, forming a distorted kagome lattice distinct from the C15-type CaNi$_2$ structure \cite{CaNi2}. This geometrical frustration can give rise to intriguing phenomena like flat bands and strong electronic correlations, potentially leading to unconventional superconducting ground states. Moreover, the high spin-orbit coupling of Re$_2$Zr could result in split electronic bands, further contributing to the formation of flat bands. The potential for these interesting features in Re$_2$Zr makes it a compelling subject for a detailed investigation of its superconducting properties.
    
\section{EXPERIMENTAL DETAILS}

A polycrystalline sample of Re$_{2}$Zr alloy was synthesized using the standard arc melting method. The phase purity of the sample was evaluated by powder X-ray diffraction (XRD) performed on a PANalytical X$^{'}$pert Pro diffractometer, equipped with Cu $K{\alpha}$ radiation ($\lambda$ = 1.54056 $\text{\AA}$). Magnetization measurements were carried out using a Superconducting Quantum Interference Device (SQUID MPMS, Quantum Design). Specific heat and electrical resistivity measurements were conducted using a Physical Property Measurement System (PPMS, Quantum Design, Inc.). Muon spin rotation and relaxation ($\mu$SR) measurements were performed using the MuSR spectrometer at the ISIS pulsed muon source at the STFC Rutherford Appleton Laboratory, United Kingdom. These measurements utilize two batches of samples that possess identical properties and characteristics.

\section{RESULTS AND DISCUSSION}

\subsection{Sample characterization}

Re$_{2}$Zr crystallizes in a hexagonal MgZn$_{2}$-type structure (C14 Laves phase), with the space group $P6_{3}$/$mmc$ (no. 194), as illustrated in \figref{Fig1}(a) using the VESTA software \cite{vesta}. Within each unit cell, two kagome layers are vertically stacked along the $c$ direction at z = 0.25 and 0.75, resulting in two non-equivalent Wyckoff sites for the Re1 and Re2 atoms. In this structure, the Re atoms form a trigonally distorted kagome lattice in the $ab$-plane, where each vertex is a convergence point for two triangles and two hexagons, as illustrated only for Re2 atoms in \figref{Fig1}(b). The difference in the Re-Re bond lengths within the Re2 layer is ($d_{Re2-Re2}$) = 2.6879 and 2.5825 \AA~ with a slight variation for the Re1 layer. These bond lengths are comparable to those of other reported breathing kagome C14 Laves phase compounds \cite{ROs2, HfOs2}, suggesting a nearly ideal 2D kagome structure with minimal variation in bond lengths within the plane. The powder XRD pattern was refined using the Rietveld method implemented in FullProf Suite software \cite{fp}, as shown in \figref{Fig1}(c). The refined lattice parameters are $a$ = $b$ = 5.270(5) \AA,  and $c$ = 8.636(2) \AA, with V$_{\text{cell}}$ = 207.782(4) \AA$^3 $, which are consistent with values previously reported in the literature \cite{Re2Zr}.

\subsection{Electrical resistivity}
\begin{figure*}
\includegraphics[width=2.05\columnwidth]{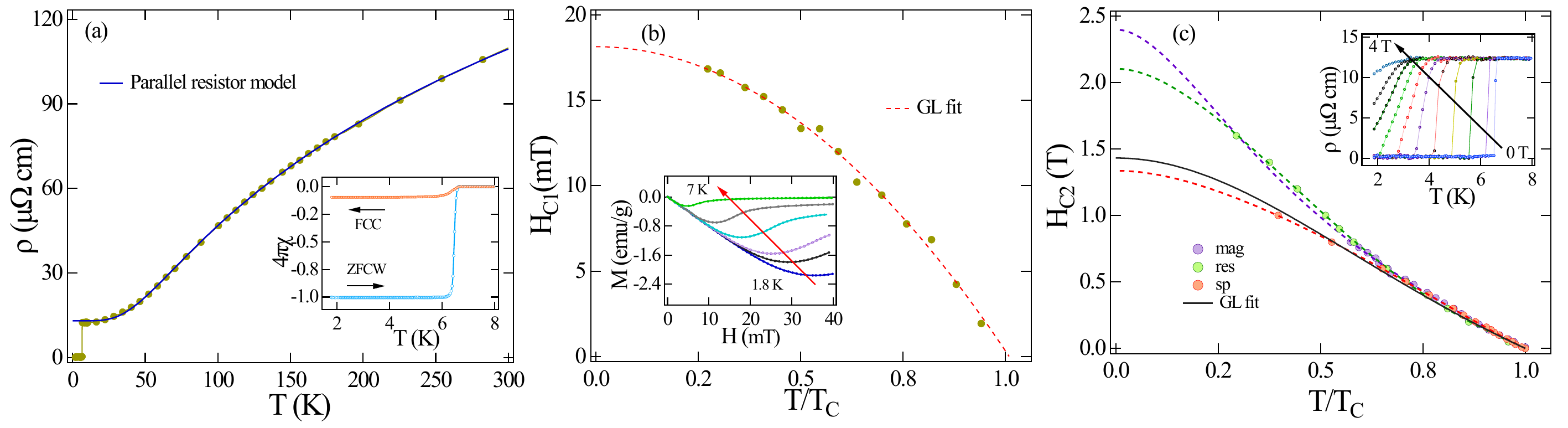}
\caption{\label{Fig2}(a) Temperature dependence of the resistivity in zero field. The inset displays magnetization data in ZFCW and FCC mode in a field of 1.0 mT. (b) Temperature dependence of the lower critical field $H_{C1}$, which was fitted using the Ginzburg-Landau relation. The inset shows low-field magnetization curves at different temperatures. (c) The zero-temperature upper critical field, $H_{C2}$(0), was determined from magnetization, resistivity, and specific heat measurements (violet circles, green circles, and orange circles, respectively). The solid black line represents the Ginzburg-Landau fit, while the two-gap fittings\cite{two_gap} are indicated by the dotted lines. The variation of $\rho$(T) under various applied magnetic fields is shown in the inset.}
\end{figure*}

The temperature dependence of the electrical resistivity $\rho$(T) in zero applied magnetic field is shown in \figref{Fig2}(a). Resistivity decreases with decreasing temperature and reaches zero at 6.65(3) K, indicating the superconducting phase transition. Additionally, magnetization measurements (inset of \figref{Fig2}(a)) reveal the presence of a superconducting state at $T_{C}^{onset}$ = 6.65(3) K (onset of a strong diamagnetic signal) with an approximate superconducting fraction of $\sim$ 100\%. The residual resistivity ratio was estimated to be $\rho$(300~K)/$\rho$(10~K) = 8.86. In the high-temperature region up to 300 K, the metallic nature of $\rho$ (T) fit well with the parallel resistor model \cite{parallel}. According to this model, the evolution of $\rho$ with temperature can be understood by the following expression:
\begin{equation}
\frac{1}{\rho(T)}=\frac{1}{\rho_{1}(T)}+\frac{1}{\rho_{sat}}
\end{equation}
where $\rho_{sat}$ is the saturation resistivity, and $\rho_{1}$(T) is the temperature-dependent resistivity. Furthermore, $\rho_{1}$(T) is given by the following expression:
\begin{equation}
\rho_{1}(T)=\rho_{0}+r\left(\frac{T}{\theta_R}\right)^{5}\int_{0}^{\theta_R/T} \frac{x^5}{(e^{x}-1)(1-e^{-x})}dx
\end{equation}
where $\rho_0$ is the residual resistivity, $\theta_{R}$ is the Debye temperature, and $r$ is a material-dependent pre-factor. This term is the Bloch-Gr\'{u}neisen term accounting for phonon-assisted electron scattering similar to s-d scattering in transition metal compounds \cite{BG}. Fitting the experimental data with this model yields $\rho_{sat}$ = 265(3) $\mu\ohm$ cm, $\theta_{R}$ = 250(3) K, and $\rho_0$ = 13.7(2) $\mu\ohm$ cm. This value of $\theta_{R}$ is close to the value obtained from the heat capacity data (described later), suggesting that the parallel resistor model can successfully explain the temperature dependence of $\rho$(T) above the superconducting region.

\subsection{Magnetization}
The lower critical field, $H_{C1}$(T), was extracted from the low-field magnetization curves $M(H)$, as shown in the inset of \figref{Fig2}(b). $H_{C1}$(0) was estimated to be 18.4(1) mT by fitting the data using the Ginzburg-Landau (GL) equation $H_{C1}(T) = H_{C1}(0)\left[1-\left(\frac{T}{T_{C}}\right)^{2}\right]$, as shown in \figref{Fig2}(b). The upper critical field, $H_{C2}$(T), was determined from magnetization, resistivity, and specific heat measurements. The inset of \figref{Fig2}(c) shows the temperature dependence of $\rho$ in the vicinity of $T_{C}$ under various applied magnetic fields. As the applied field increased, $T_{C}$ shifted to a lower value, as expected for a superconductor. In the high-field and low-temperature region, the data deviate from the GL relation $H_{C2}(T) = H_{C2}(0)\frac{(1-t^{2})}{(1+t^2)}$, where $t = T/T_{C}$ (\figref{Fig2}(c)). The data were well fitted with the two-gap model \cite{two_gap, Re2Hf}, as indicated by the dotted lines in \figref{Fig2}(c). From the magnetization data, the estimated value of $H_{C2}^{mag}(0)$ is 2.39(1) T. The Pauli limiting field within the BCS theory is given by $H_{C2}^{p}$(0) = 1.83 T$_{C}$, which gives $H_{C2}^{p}$(0) = 12.0(1) T.


The Ginzburg-Landau coherence length $\xi_{GL}(0)$ was calculated to be 117(2) \text{\AA} using the relation $H_{C2}(0)$ = $\frac{\Phi_{0}}{2\pi\xi_{GL}^{2}}$, where $\Phi_{0}$ is the magnetic flux quantum (= 2.07 $\times$10$^{-15}$ T m$^{2}$) \cite{tin}. The GL penetration depth, $\lambda_{GL}$(0), was subsequently determined to be 1515(3) $\text{\AA}$, using the relations Eq.\eqref{eqn8_5:PEN},
\begin{equation}
H_{C1}(0) = \left(\frac{\Phi_{0}}{4\pi\lambda_{GL}^2(0)}\right)\ln\left(\frac{\lambda_{GL}(0)}{\xi_{GL}(0)} + 0.497\right). 
\label{eqn8_5:PEN}
\end{equation}    
The GL parameter $k_{GL} = \frac{\lambda_{GL}(0)}{\xi_{GL}(0)}$ was found to be 12.9(2), which is much higher than $\frac{1}{\sqrt{2}}$, and classifies Re$_{2}$Zr as a type-II superconductor. The thermodynamic critical field, $H_{C}$, is related to the lower and upper critical fields by the relation $ H_{C1}(0)H_{C2}(0) = H_{C}^2 \ln k_{GL}$, with a calculated value of $H_{C}$ around 0.13(2) T.

\subsection{Specific heat}
\begin{figure*}
\includegraphics[width=2.05\columnwidth]{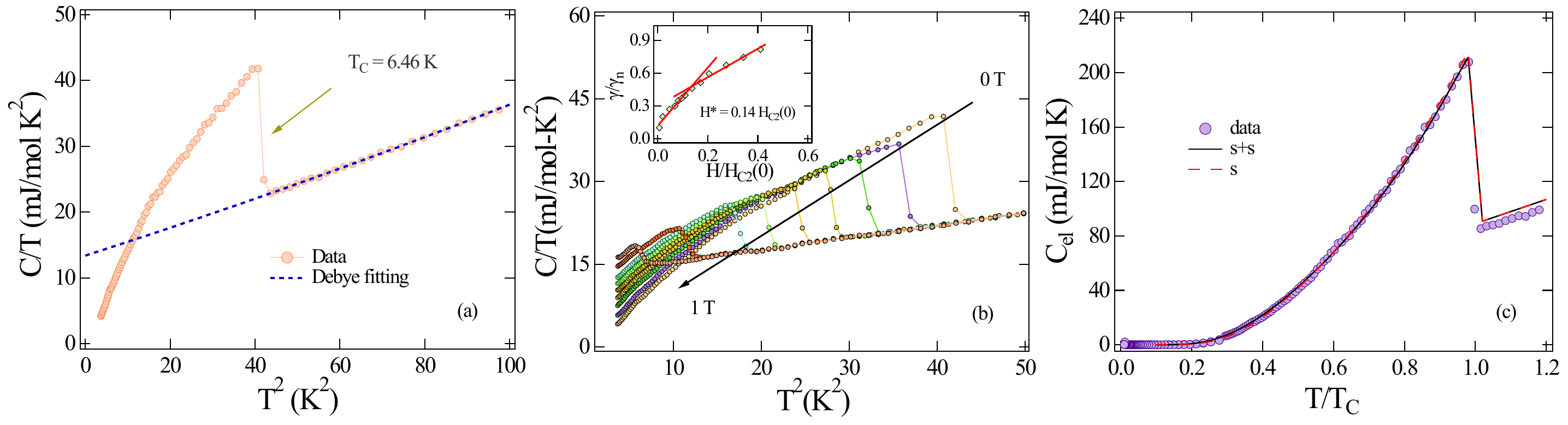}
\caption{\label{Fig3}Specific heat divided by temperature, $C(T)/T$, plotted against $T^{2}$ in (a) zero field fitted with the Debye model and (b) as a function of applied magnetic field. The variation of the Sommerfeld coefficient $\gamma$ with respect to the magnetic field is depicted in the inset. (c) The temperature dependence of $C_{el}$ fitted using the $s$ (red dotted line) and $s+s$ models (black line) using Eq.\eqref{two_gap}.}
\end{figure*}

Temperature-dependent specific heat measurements, $C(T)$, were performed to analyze the superconducting state of the sample. The zero field $C(T)/T$ as a function of T$^{2}$ is shown in \figref{Fig3}(a), while the magnetic field dependence is presented in \figref{Fig3}(b) over the range of 0-1 T. A pronounced jump in the specific heat data at 6.46(2) K confirms bulk superconductivity. The low-temperature normal-state specific heat data were fitted using the relation,
\begin{equation}  
\frac{C}{T}=\gamma_{n}+\beta_{3}T^{2}+\beta_{5}T^{4}  
\label{eqn3:hc}    
\end{equation}  
where $\gamma_{n}$ is the Sommerfeld coefficient, $\beta_{3}$ is the Debye constant, and $\beta_{5}$ is the anharmonic contribution to the specific heat. We obtain the fitting parameters as $\gamma_{n}$ = 12.9(4) mJ mol$^{-1}$ K$^{-2}$, $\beta_{3}$ = 0.22(1) mJ mol$^{-1}$ K$^{-4}$ and $\beta_{5}$ = 1.1(7)$\times$10$^{-4}$ $\mu$J mol$^{-1}$ K$^{-6}$. From the experimental value of $\beta_{3}$, the Debye temperature $\theta_{D}$ of the compound can be calculated using the simple Debye model for the phonon contribution to the specific heat in \equref{eqn4:dt}:
   \begin{equation} 
   \theta_{D}= \left(\frac{12\pi^{4}RN}{5\beta_{3}}\right)^{\frac{1}{3}},
    \label{eqn4:dt}  
    \end{equation} 
where $N$ (= 3) is the number of atoms per formula unit and $R$ is the molar gas constant (= 8.314 J mol$^{-1}$ K$^{-1}$). The estimated value of $\theta_{D}$ is 297(5) K. For non-interacting particles, $\gamma_{n}$ is proportional to the density of states $D_{C}(E_{\mathrm{F}})$ at the Fermi level, which is calculated to be 5.47 $\frac{states}{eV f.u.}$ from the relation $\gamma_{n}= \left(\frac{\pi^{2}k_{B}^{2}}{3}\right)D_{C}(E_{\mathrm{F}})$, where $k_{B}$ $\approx$ 1.38 $\times$ 10$^{-23}$ J K$^{-1}$. The electron-phonon coupling constant $\lambda_{e-ph}$ is estimated as 0.70(1) using the McMillan theory \cite{Mc1}, stated as: 
 \begin{equation}
 \lambda_{e-ph} = \frac{1.04+\mu^{*} ln(\theta_{D}/1.45T_{c})}{(1-0.62\mu^{*})ln(\theta_{D}/1.45T_{c})-1.04 },
 \label{eqn8:ld}
 \end{equation}  
where $\mu^{*}$ (= 0.13 for many intermetallic superconductors) is the repulsive screened Coulomb parameter. This value of $\lambda_{e-ph}$ suggests that Re$_{2}$Zr is a moderately coupled superconductor similar to Re$_{2}$Hf \cite{Re2Hf}, Zr$_{2}$Ir \cite{Zr2Ir} and other A15 compounds such as Ti$_{3}$Ir \cite{A15} and Ti$_{3}$Sb \cite{A15}.

The electronic contribution to the specific heat in the superconducting state $C_{el}(T)$ can be calculated by subtracting the phononic contribution from the measured data $C(T)$ using the equation $C_{el}(T) = C(T)-\beta_{3}T^{3}-\beta_{5}T^{5}$. The magnitude of the specific heat jump $\frac{\Delta C_{el}}{T_{C}}$ at $T_{C}$ is 19.33 mJ mol$^{-1}$ K$^{-2}$. This gives a normalized specific heat jump of $\frac{\Delta C_{el}}{\gamma_{n}T_{C}}$ = 1.50(5), which is slightly higher than the BCS value (1.43) in the weak coupling limit. In the superconducting state, the Sommerfeld coefficient $\gamma$ was calculated by fitting $C_{el}$/T versus $T$ with equation $\frac{C_{el}}{T}$ = $\gamma + \frac{A}{T}\exp\left(\frac{-bT_{C}}{T}\right)$, with $A$ and $b$ being the fitting parameters \cite{sas}. The field dependence of $\gamma$ is shown in the inset of \figref{Fig3}(b), where $\gamma$ and $H$ are normalized by $\gamma_{n}$ and $H_{C2}$(0). Here we observe that in the lower field region $\gamma$ increases linearly as a function of $H$ up to the crossover field $H^{*} = 0.14 \times H_{C2}$ (0), which is slightly less than the proposed theoretical value of $0.3 \times H_{C2}$(0) for a completely isotropic gap superconductor \cite{line_nodes,line_nodes2}. Similar behaviour has been observed in Re$_{2}$Hf \cite{Re2Hf}. The field dependence observed in $\gamma$ suggests the possibility of an unconventional superconducting energy gap. To find the nature of the energy gap, the electronic contribution to the specific heat data at zero field in the superconducting region was fitted using the $\alpha$-model \cite{alpha} as follows:
\begin{equation}
\frac{C_{el}}{\gamma_{n}T_{C}} = t\frac{d(S/\gamma_{n}T_{C})}{dt},
\end{equation}
\begin{equation}
\frac{S}{\gamma_{n}T_{C}}(t) = -\frac{3}{\pi^3}\int_{0}^{2\pi}\int_{0}^{\infty}[f \ln f + (1-f) \ln (1-f)] d\epsilon d\theta,
\end{equation}
\begin{equation}
f = \left[\exp \left(\frac{(\epsilon^{2}+A^{2}(t,\theta)/4)^{0.5}}{t}\right)+1\right]^{-1},
\end{equation}
and
\begin{equation}
\delta(T) = \frac{\Delta(T)}{\Delta(0)} = \tanh\left[1.82\left(1.018\left(\frac{T_C}{T} - 1\right)\right)^{0.51}\right],
\end{equation}
where $t = \frac{T}{T_{C}}$, $A(t, \theta)$= $\alpha $g$_{k}$($\theta$)$\delta(T)$, and $\alpha$ = 2$\frac{\Delta(0)}{k_{B}T_{C}}$. g$_{k}$($\theta$) is the azimuthal angle-dependent part of the energy, which is different for different gap functions and depends on the gap symmetry. For an isotropic $s$ gap, g$_k$ is taken as 1. In the case of the two-band phenomenological $\alpha$-model, the contributions of each band can be calculated within the above model with corresponding weight factors $\omega$ and (1 - $\omega$) as follows:
\begin{equation}
C(T) = \omega C_{1}(T) + (1 - \omega) C_{2}(T)
\label{two_gap}
\end{equation}

\begin{figure*}
\includegraphics[width=2.05\columnwidth]{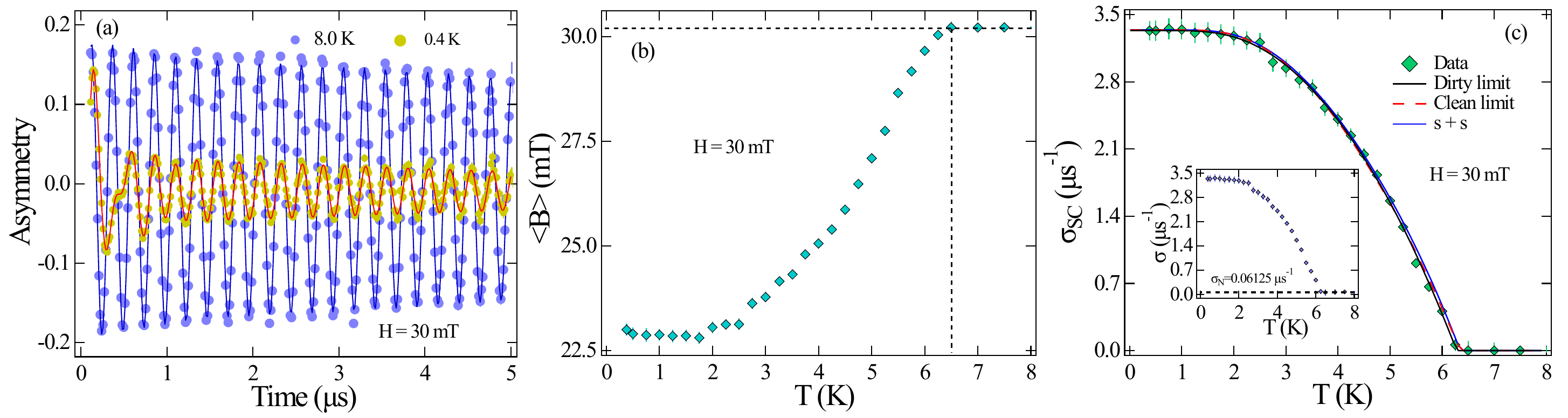}
\caption{\label{Fig4} (a) TF-$\mu$SR asymmetry spectra measured above ($T$ = 8.0 K) and below ($T$ = 0.4 K) the $T_{C}$ in an applied field $H$ = 30 mT, where solid lines represent fits the data using Eq.~$\ref{eqn1_5}$. (b) Temperature variation of the internal magnetic field $\langle B \rangle$. (c) Temperature dependence of the depolarization rate $\sigma_{sc}$ is fitted using a dirty and clean limit $s$-wave model, represented by the black solid line and red dotted line, respectively, along with the  $s+s$ model (blue solid line). Inset shows the temperature dependence of the effective depolarization rate $\sigma$.}
\end{figure*}
 
The best fit to the data was obtained with the $s+s$ model compared to the $s$ model, as shown in \figref{Fig3}(c).
The gap values obtained by fitting $C_{el}$ with a two-gap model are 1.9 and 1.4, with a weight factor (see the fraction of the gap, $\omega$) of 0.68 (goodness of fit, \(\chi^2 = 1.21\)). In comparison, the gap value for a single $s$-wave model is $\frac{\Delta(0)}{k_{B}T_{C}} = 1.7$ (\(\chi^2 = 1.35\)). To find the exact superconducting gap structure, local probe measurements, e.g., muon spin rotation and relaxation measurements, are vital.

    
\subsection{Muon Spin Rotation and Relaxation}
Transverse field (TF) $\mu$SR measurements were performed in the vortex state to investigate the superconducting gap structure. Measurements were performed after the sample was field-cooled, where a field $H$ = 30 mT ($H_{C1} < H < H_{C2}$) was applied perpendicular to the initial muon spin polarization. The TF-$\mu$SR spectra collected above and below $T_{C}$ are shown in \figref{Fig4}(a). The normal state ($T$ = 8.0 K) shows oscillatory spectra suggesting a homogeneous field distribution throughout the sample, with weak depolarization arising from the dipolar nuclear field. In contrast, the asymmetry spectra in the superconducting state ($T$ = 0.4 K) show strong depolarization, indicating an inhomogeneous field distribution resulting from the flux line lattice state. The time-domain spectra were best fitted using a multicomponent Gaussian damped oscillatory function with an undamped oscillatory background term that emerges from the muons implanted directly into the silver sample holder that does not depolarize, represented as:
\begin{eqnarray}
G_{\mathrm{TF}}(t) &=&
\sum_{\mathrm{i}=1}^{N} A_{\mathrm{i}}\mathrm{exp}\left(\frac{-\sigma_{\mathrm{i}}^{2}t^{2}}{2}\right)\mathrm{cos}(\gamma_{\mu}B_{\mathrm{i}}t+\phi)\nonumber\\&+&A_{bg}\mathrm{cos}(\gamma_{\mu}B_{bg}t+\phi)
\label{eqn1_5}
\end{eqnarray}
where $\phi$ is the phase of the initial muon spin polarization with respect to the positron detector. $A_{\mathrm{i}}$ and $B_{\mathrm{i}}$, are the asymmetry and mean field (first moment) of the $\mathrm{i}^{th}$ component of the Gaussian distribution, respectively. $\sigma_{\mathrm{i}}$ is the depolarization/relaxation rate and $\gamma_{\mu}$/2$\pi$ = 135.5 MHz/T is the muon gyromagnetic ratio. $A_{bg}$ and $B_{bg}$ are the background contributions for the asymmetry and the field, respectively. The temperature dependence of the effective depolarization rate $\sigma$ was calculated using the second-moment method \cite{khasanov2005magnetic}. Here, the first and second moments are described as Eq.\eqref{eqn2_ch3} and Eq.\eqref{eqn3_ch3}:
    \begin{eqnarray}
   \langle B \rangle &=& \sum_{\mathrm{i}=1}^{N}\frac{A_{\mathrm{i}}\mathrm{B_{i}}}{A_{1}+A_{2}+.....+A_{\mathrm{N}}}
    \label{eqn2_ch3}
    \end{eqnarray}
    
    \begin{eqnarray}
    \langle \Delta B^{2} \rangle = \sum_{\mathrm{i}=1}^{\mathrm{N}}\frac{A_{\mathrm{i}}[(\sigma_{\mathrm{i}}/\gamma_{\mu})^2+(B_{\mathrm{i}}-\langle B \rangle)^2]}{A_{\mathrm{1}}+A_{\mathrm{2}}+.....+A_{\mathrm{N}}}= \frac{\sigma^2}{\gamma_{\mu}^2}.
    \label{eqn3_ch3}
    \end{eqnarray}
    
The asymmetry spectra (\figref{Fig4}(a)) were fitted with three Gaussian components. The temperature variation of the internal magnetic field, $\langle B \rangle$, is depicted in \figref{Fig4}(b), while the temperature dependence of the effective depolarization rate $\sigma$, as extracted using Equations (\ref{eqn1_5}-\ref{eqn3_ch3}), is shown in the inset of \figref{Fig4}(c). $\sigma$ incorporates depolarization arising from the nuclear dipole moments ($\sigma_{\mathrm{N}}$) and the field variation across the flux line lattice ($\sigma_{\mathrm{sc}}$), as expressed by the quadratic relation: $\sigma^{2}$ = $\sigma_{\mathrm{sc}}^{2}+\sigma_{\mathrm{N}}^{2}$, which we used to extract $\sigma_{\mathrm{sc}}$ from the data.

\begin{figure*}
\includegraphics[width=2.08\columnwidth]{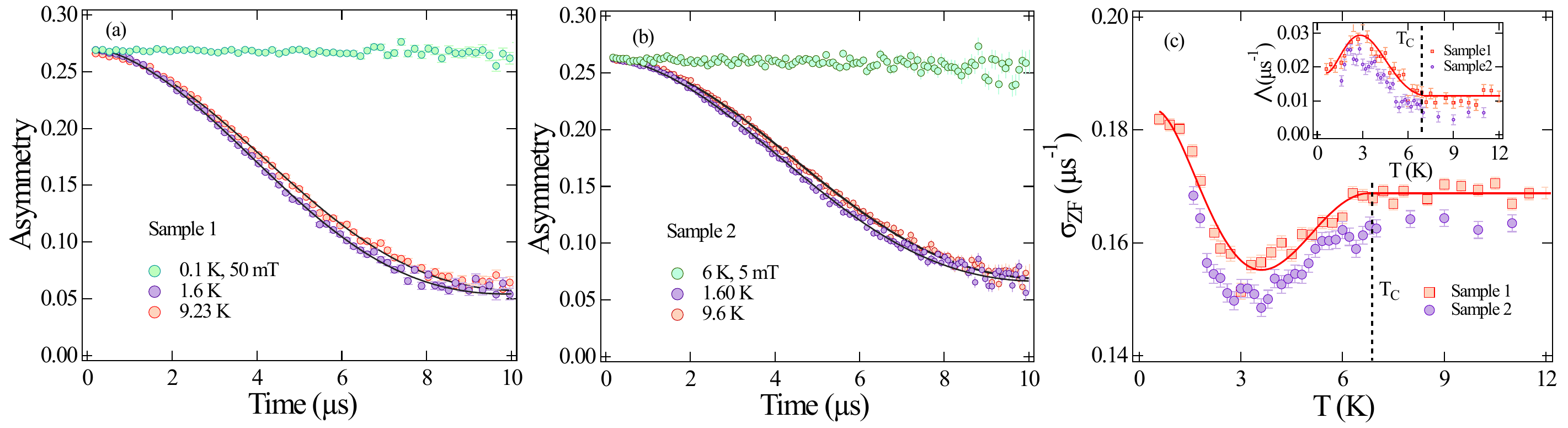}
\caption{\label{Fig5} (a) and (b) Time evolution of the spin polarization of muons above and below $T_{C}$ under zero-field conditions for two batches of sample. The solid lines show the fitting of the asymmetry spectra using  Eq.~$\ref{zf_eq}$ and $\ref{zf_eq1}$. (c) Temperature dependence of relaxation rate $\sigma_{\mathrm{ZF}}$. The inset shows the temperature dependence of relaxation rate $\Lambda$. Here, the solid red line represents a guide-to-the-eye for the change in relaxation rates.}
\end{figure*}

The temperature dependence of $\sigma_{\mathrm{sc}}$ (\figref{Fig4}(c)) is seen to be nearly constant below $\approx$ $T_{C}$/3, indicating the absence of nodes in the superconducting energy gap at the Fermi surface. This nature can be well described by the s-wave model in the dirty limit as given by;
\begin{equation}  
\frac{\sigma_{sc(T)}}{\sigma_{sc}(0)} = \frac{\lambda^{-2}(T)}{\lambda^{-2}(0)} = \frac{\Delta_0(T)}{\Delta_0(0)}\tanh\left[\frac{\Delta_0(T)}{2k_{B}T}\right]
\label{eqn5_3}
\end{equation}
$\Delta_{0}(T)$/$\Delta_{0}(0
)$ = $\tanh\{1.82(1.018({T_{C}/T}-1))^{0.51}\}$ is the BCS approximation for temperature-dependent energy gap and $\Delta_{0}(0)$ is the gap value at zero temperature. We have also fitted the data within the semiclassical approximation of the clean limit using Eq.\eqref{eqn5_3}:
\begin{equation}  
\frac{\sigma_{sc}^{-2}(T,\Delta_{0})}{\sigma_{sc}^{-2}(0,\Delta_{0})} = 1+\frac{1}{\pi}\int_{0}^{2\pi}\int_{\Delta_k(T,\phi)}^{\infty}\left(\frac{\delta f}{\delta E }\right)\frac{E dE  d\phi}{\sqrt{E^{2}-\Delta_{k}^{2}(T,\phi)}},
\label{eqn5_3}
\end{equation}
where $ f $ = [1 + exp(E/k$_{B}T)]^{-1}$ is the Fermi function and $\Delta_{k}(T,\phi)$= $\Delta_{0}(T)$g$_{k}(\phi)$. $\Delta_{0}(T)$ is the value of the BCS gap and $g_{k}$ is the angular dependence of the gap, which depends on the gap symmetries \cite{khasanov2008evidence}. Moreover, for the two-gap model, the temperature dependence of $\sigma_{sc}^{-2}$ can be expressed as the linear combination of two components, Eq.\eqref{clean:s_3}:
\begin{equation}
\frac{\sigma_{sc}^{-2}(T)}{\sigma_{sc}^{-2}(0)} = \omega \frac{\sigma_{sc}^{-2}(T, \Delta_{0,1})}{\sigma_{sc}^{-2}(0,\Delta_{0,1})} + (1 - \omega) \frac{\sigma_{sc}^{-2}(T, \Delta_{0,2})}{\sigma_{sc}^{-2}(0,\Delta_{0,2})},
\label{clean:s_3}
\end{equation}

where $\Delta_{0,1}$ and $\Delta_{0,2}$ are the gap waves with the weight factor $\omega$ and ($1- \omega$). We have obtained a good fit for the data using the s-wave dirty limit model, providing a gap value of $\frac{\Delta_0(0)}{k_{B}T_{C}}$ = 1.65(1) (\(\chi^2 = 1.12\)). The $\mu$SR data fitted with the $s+s$ model yield gap values of 2.35 and 1.83, with a weight factor of 0.25 (\(\chi^2 = 1.30\)), while for the single s-wave clean-limit fit, the gap value is $\frac{\Delta(0)}{k_{B}T_{C}}$ = 1.90 (\(\chi^2 = 1.24\)). The different model fittings for the temperature dependence of $\sigma_{sc}$ are shown in \figref{Fig4}(c).
In a type-II isotropic superconductor with a hexagonal Abrikosov vortex lattice having $\kappa$ $\mathrm{>}$ 5, the penetration depth $\lambda$ can be calculated with a high degree of accuracy, using Eq. \eqref{clean:s2_3} \cite{lambda1,lambda2},
\begin{equation}
\sigma_{\mathrm{sc}} [\mu s^{-1}] = 4.854\times 10^{4}(1-h)[1+1.21(1-\sqrt{h})^{3}]\lambda^{-2} [nm^{-2}],
\label{clean:s2_3}
\end{equation} 
where $h$ = $ H/H_{C2}(T) $ is the reduced field. Using the measured $H_{C2}$ (T), $\lambda$ was determined to be 1628(2) $\text{\AA}$.\\

The asymmetry spectra in zero field (ZF), measured above and below the superconducting transition temperature, are shown in \figref{Fig5}(a) and (b) for two batches of samples labeled 1 and 2. Both batches were synthesized using the same preparation method and displayed similar superconducting characteristics based on magnetization and specific heat measurements. A slight change in relaxation behavior is observed in the superconducting state compared to the normal state for both batches. Furthermore, a low longitudinal field of 50 mT in the superconducting state (0.1 K) fully decouples the muon spins from relaxation, as shown in \figref{Fig5} (a), while a similar decoupling effect is observed for batch 2 at 6 K and 5 mT.
This suggests that the magnetic field in the superconducting state is quasi-static for both batches. To better understand this behavior, ZF-$\mu$SR data were fitted using a damped Gaussian Kubo-Toyabe function \cite{PhysRevB.20.850} with a background contribution associated with muons stopping in the silver sample holder. The asymmetry fitting function is given by:
\begin{equation}
G(t)=A_{1}\mathrm{exp}(-\Lambda t)G_{\mathrm{KT}}(t)+A_{\mathrm{bg}} ,
\label{zf_eq}
\end{equation}
with \begin{eqnarray}
G_{\mathrm{KT}}(t) &=& \frac{1}{3}+\frac{2}{3}(1-\sigma^{2}_{\mathrm{ZF}}t^{2})\mathrm{exp}\left(\frac{-\sigma^{2}_{\mathrm{ZF}}t^{2}}{2}\right),
\label{zf_eq1}
\end{eqnarray}
where $A_{1}$ is the initial sample asymmetry, and $\sigma_{\mathrm{ZF}}$ and $\Lambda$ are the Gaussian and an additional relaxation rate, respectively. All parameters were allowed to vary freely during the fitting of the asymmetry spectra, and the temperature dependence of the relaxation rates for the two samples is shown in \figref{Fig5}(c). The background and sample asymmetry, $A_{bg}$ and $A_1$ are almost temperature independent with no observable trend, while relaxation rates $\sigma_{\mathrm{ZF}}$ and $\Lambda$ show a significant change in the superconducting state. Above T$_C$ (marked by a vertical dashed black line), both relaxation parameters $\sigma_{\mathrm{ZF}}$ and $\Lambda$ are temperature independent. In the superconducting state, as the temperature decreases, $\sigma_{\mathrm{ZF}}$ initially decreases until it reaches a temperature T$^{'}$ $\approx$ 3.1 K, where it starts to increase again but does not saturate to the lowest measured temperature, creating a valley-like behavior at T$^{'}$, a temperature significantly different from T$_C$. In contrast, $\Lambda$ exhibits a peak-like behaviour at the same temperature T$^{'}$. The increase in relaxation $\sigma_{sc}$ below T$^{'}$ can be associated with the presence of a spontaneous magnetic field, indicating the break of the time-reversal symmetry in the superconducting state. For sample 1, the calculated spontaneous field $B_{int}$ = 0.22 G is comparable to that observed in other time-reversal symmetry-broken superconductors, providing strong evidence for a broken TRS pairing state\cite{Re2Hf, prpt4ge12, hillier2012nonunitary, Re}.


We explored various approaches to understand this phenomenon. One possibility is that the two relaxation rates, $\sigma_{\mathrm{ZF}}$ and $\Lambda$, may be interdependent, causing one to exhibit a valley and the other a peak. To test this hypothesis, we performed fits while keeping $\Lambda$ fixed. A similar valley trend with significant change is observed in $\sigma_{\mathrm{ZF}}$; however, this approach resulted in a poor fit of the asymmetry spectra. We further attempted to fit the spectra using various fitting functions, including the stretched and Lorentzian-damped Gaussian Kubo-Toyabe functions. However, a valley trend in the relaxation parameters persisted. Notably, this behavior was consistently observed in both sample sets, 1 and 2, measured in different sample environments and a year apart, underscoring its intrinsic nature and independence from external parameters. Additionally, changes in both the channels of the muon relaxation are not new and have previously been observed in several TRS broken systems such as La$_7$Ir$_3$, La$_7$Pd$_3$, La$_7$Rh$_3$ \cite{La7Ir3,la7pd3, La7rh3} and Pr$_{1-x}$La$_{x}$Pt$_4$Ge$_{12}$, Pr(Os$_{1-x}$Ru$_x$)$_4$Sb$_{12}$, Pr$_{1-y}$La$_y$Os$_4$Sb$_{12}$ \cite{Prrusb12,Prlaptge12} and Lu$_3$Os$_4$Ge$_{13}$ \cite{lu3os4ge13}. In these cases, the increase in the secondary relaxation channel is attributed to nuclear spin fluctuations, which may also be the cause of the observed increase starting at T$^{'}$. In particular, our magnetization, specific heat, and resistivity measurements, along with TF-$\mu$SR results, do not exhibit any anomaly at T$^{'}$, and longitudinal field $\mu$SR measurements also rule out the presence of any external magnetic field impurity. The valley/dip in the relaxation rate is observed in many other TRS broken superconductors PrPt$_4$Ge$_{12}$ \cite{prpt4ge12}, Pr$_{1-x}$Ce$_{x}$Pt$_4$Ge$_{12}$ \cite{Prceptge12}, Pr$_{1-x}$La$_{x}$Pt$_4$Ge$_{12}$ \cite{Prlaptge12} and recently in La$_7$Ni$_3$ \cite{la7ni3} and Lu$_3$Os$_4$Ge$_{13}$ \cite{lu3os4ge13}. These results suggest the possibility of multicomponent order parameters with a multi-gap superconducting state.  

It is also worth noting that while this trend has been observed in some potential multicomponent, multigap superconductors, it is not specific to any crystal symmetry or spin-orbit coupling (SOC) strength. For example, La$_7$Ni$_3$ shows a dip feature, whereas La$_7$Ir$_3$, La$_7$Pd$_3$, and La$_7$Rh$_3$ do not, where La$_7$Ni$_3$ has lower SOC compared to others \cite{la7ni3,La7rh3, La7Ir3,la7pd3}. Similarly, Lu$_3$Os$_4$Ge$_{13}$ with high SOC exhibits a dip feature in the relaxation, whereas Y$_3$Ru$_4$Ge$_{13}$ does not \cite{lu3os4ge13, y3ru4ge13}. In our case, the dip/valley-like trend is noted in Re$_2$Zr but not in Re$_2$Hf \cite{Re2Hf}, suggesting that the underlying mechanism may be complex and potentially related to the interplay of multiple superconducting gaps within the material.

Theoretical studies on Re$_2$Hf have suggested that geometric frustration-driven spin-fluctuations could mediate unconventional superconducting pairing. Given the similarity between Re$_2$Zr and Re$_2$Hf, and the evidence of spin-fluctuations in Re$_2$Zr from $\mu$SR measurements, it is plausible that this mechanism might also be at play in Re$_2$Zr\cite{Re2Hf}. While Re$_2$Zr and Re$_2$Hf differ in their spin-orbit coupling due to the 4d/5d elements, the presence of geometric frustration within the kagome layer appears to be a crucial factor in establishing an unconventional superconducting ground state in both compounds. Moreover, the Re element itself might be responsible for the unconventional superconducting ground state\cite{Re}. However, the complex unit cell structure of these compounds\cite{ROs2, HfOs2}, comprising two kagome layers and a triangular layer, makes it challenging to definitively pinpoint the pairing mechanism. To achieve a conclusive understanding, further investigations involving single-crystal experiments and in-depth band structure calculations are necessary.

\begin{figure} 
    \includegraphics[width=1.0\columnwidth]{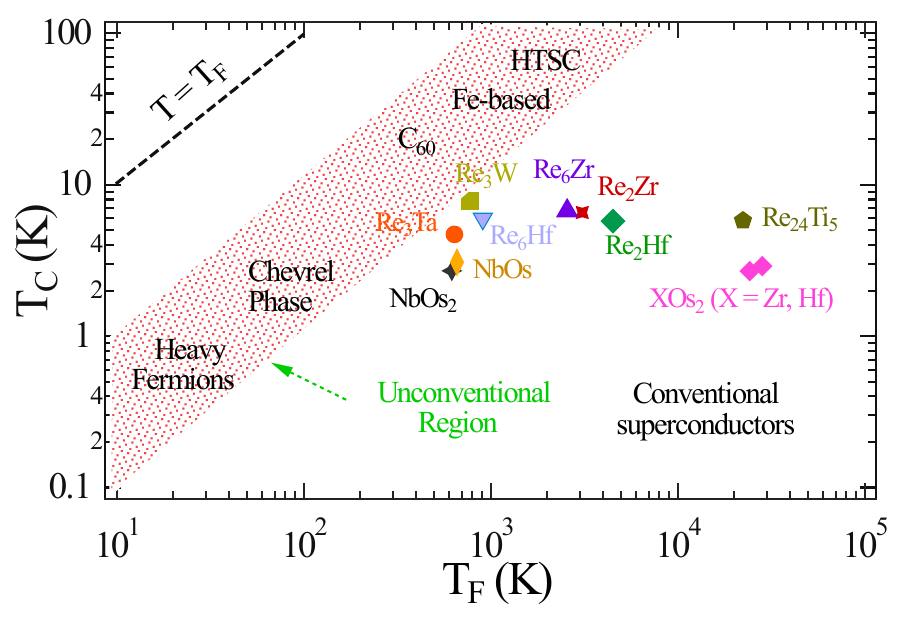}
    \caption{\label{Fig6} Uemura plot of Re$_{2}$Zr along with Re$_{2}$Hf \cite{Re2Hf} and Re$_{6}$X \cite{ReZr, Re6Ti, Re24Ti5}, and other unconventional materials represented by the red band.}
\end{figure}

Uemura et al. \cite{uemura1,uemura2,uemura3,uemura4} have described a method for classifying superconductors based on the ratio of the critical temperature $T_{C}$ to the effective Fermi temperature $T_{\mathrm{F}}$. In this classification, exotic superconductors fall in the range 0.01 $\leq \frac{T_C}{T_{\mathrm{F}}} \leq $ 0.1, whereas conventional BCS superconductors have $\frac{T_{C}}{T_\mathrm{F}} \leq $ 0.001. We calculated \(T_{\mathrm{F}}\) using the relation \( k_B T_{\mathrm{F}} = \frac{\hbar^2}{2} \left(3\pi^2\right)^{2/3} \frac{n_s^{2/3}}{m^*} \) \cite{PhysRevB.102.094515}. Here, we estimate the superconducting carrier density \( n_s \) by \( n_s(0) = \frac{m^*}{\mu_0 e^2 \lambda^2} \), where \( m^* = (1 + \lambda_{e-\text{ph}})m_e \). Using \(\lambda = 1628 \, \text{\AA}\) (from muon spectroscopy measurements) and \(\lambda_{e-\text{ph}} = 0.70\) (from specific heat data), we estimated \(T_{\mathrm{F}}\) = 3091(6) K, resulting in a corresponding value of \(\frac{T_C}{T_{\mathrm{F}}} = 0.0021(1)\) for Re$_2$Zr. This ratio places the compound Re$_2$Zr outside the band of unconventional families but close to the broken TRS superconductors Re$_{2}$Hf \cite{Re2Hf} and Re$_6$X \cite{ReZr, Re6Ti, Re24Ti5}. The Uemura plot of Re$_2$Zr, along with Re$_2$Hf and some other unconventional materials \cite{NbOs, NbOs2, uemura5, uemura6, uemura7, uemura8}, is shown in \figref{Fig6}.  A summary of all experimentally measured and estimated parameters, compared with those for Re$_2$Hf and Re$_6$Zr, is provided in \tableref{tbl_3}.

  
    \begin{table}
    \caption{Normal and superconducting parameters of Re$_{2}$Zr in comparison with Re$_{2}$Hf \cite{Re2Hf} and Re$_{6}$Zr \cite{ReZr}.}
    \label{tbl_3}
    \setlength{\tabcolsep}{5pt}
    \begin{center}
    \begin{tabular}[b]{|l c c c c| }\hline 
    Parameters& unit& Re$_{2}$Zr & Re$_{2}$Hf  &  Re$_{6}$Zr \\
    \hline
    T$_{C}$ & K &  6.6(2)  & 5.7(2) & 6.75 \\
    $\rho_{0}$ & $ \mu\Omega $cm & 13.7(2)  & 22.0(1) &    142 \\
    H$_{C1}(0)$& mT & 18.4(2)  &12.7(2) &    10.3\\ 
    H$_{C2}$(0) & T&  2.39(1)  & 1.17(2) &  11.2\\
    H$_{C2}^{P}$(0) & T &  12.0(1)  &10.5 &    12.35\\
    $\xi_{GL}$& $\text{\AA}$ & 117(2) &168 &     53.7\\
    $\lambda_{GL}$& $\text{\AA}$ & 1515(3)  &1739 &    2470\\
    $k_{GL}$& & 12.9(2) &10.36(1) &    46.2\\
    $\gamma_{n}$& mJ/mol K$^{2}$ & 12.9(4)  &11.04(2) &    26.9\\
    $\theta_{D}$& K & 297(5) & 293.8 &  338\\
    $\frac{\Delta C_{el}}{\gamma_{n}T_{C}}$ &   & 1.50(5) &1.41 &     1.60\\
    $\lambda_{e-ph}$ &  & 0.70(1)  &0.67 &    0.67\\ 
    D$_{C}$(E$_{\mathrm{F}}$)& states/eV f.u.& 5.47 &4.69 &  \\
    n & 10$^{26}$m$^{-3}$ & 18.0  &24.8 &  15.2\\
    $\frac{m*}{m_{e}}$ &  & 1.70  &7.8 &   10.1 \\
    $T_{\mathrm{F}}$& K & 3091(6) &4514 &  2570 \\
    $\frac{T_{C}}{T_{\mathrm{F}}}$& & 0.0021(1)  &0.0013 &  0.0026\\    
    \hline
    \end{tabular}
    \par\medskip\footnotesize
    \end{center}
    \end{table} 

\section{Conclusion}
In summary, we have investigated the Re-based kagome lattice superconductor Re$_2$Zr through transport, magnetization, specific heat, and muon-spin rotation and relaxation measurements. Our findings unequivocally establish Re$_2$Zr as a type-II superconductor with moderately coupled Cooper pairs. The upward curvature observed in the upper critical field and the specific heat data suggest that Re$_2$Zr may exhibit characteristics of a two-nodeless gap superconductor akin to the behavior observed in the Re$_2$Hf alloy. Furthermore, the non-linear response of $\gamma$ at low fields hints at the presence of an unconventional superconducting energy gap. The transverse field muon-spin rotation (TF-$\mu$SR) data further confirm the existence of two-gap superconductivity. Furthermore, zero-field and longitudinal-field $\mu$SR data suggest time-reversal symmetry breaking in the superconducting ground state and suggest a potential unconventional pairing mechanism, which underlies the superconductivity in Re$_2$Zr. This comprehensive investigation underscores the need for further exploration, particularly on Re$_2$Zr single crystals, to elucidate potential complex-order parameters similar to those observed in Re$_{2}$Hf. This work will be valuable for understanding the complex superconducting ground state of kagome-structured superconductors, where the pairing mechanism remains elusive.

\section{Acknowledgments}R.P.S. acknowledges the SERB Government of India for the Core Research Grant No. CRG/2023/000817.

\bibliographystyle{apsrev4-2}
\bibliography{references} 

\end{document}